\documentclass[aps,showpacs,nofootinbib,superscriptaddress,eqsecnum]{revtex4}
\usepackage[mathscr]{eucal}
\usepackage{color}
\usepackage[dvips]{graphicx}
\usepackage{epsf}
\usepackage{bm}
\usepackage{epsfig}
\usepackage{feynmf}
\usepackage{amsbsy,amsmath,amsfonts}

\newcommand{\be}{\begin{equation}}
\newcommand{\ee}{\end{equation}}
\newcommand{\ba}{\begin{eqnarray}}
\newcommand{\ea}{\end{eqnarray}}

\newcommand{\pfe}{p_{e}}
\newcommand{\pfp}{p_{p}}

\begin{document}

\title{ Shear viscosity and the r-mode instability window in superfluid neutron stars}

\author{Cristina Manuel}
\email{cmanuel@ieec.uab.es}
\affiliation{Instituto de Ciencias del Espacio (IEEC/CSIC) Campus Universitat Aut\`onoma de Barcelona, Facultat de Ci\`encies, Torre C5, E-08193 Bellaterra (Barcelona), Catalonia, Spain}

\author{Laura Tolos}
\email{tolos@ice.csic.es}
\affiliation{Instituto de Ciencias del Espacio (IEEC/CSIC) Campus Universitat Aut\`onoma de Barcelona, Facultat de Ci\`encies, Torre C5, E-08193 Bellaterra (Barcelona), Catalonia, Spain}
\affiliation{Frankfurt Institute for Advanced Studies, Johann Wolfgang Goethe University,
Ruth-Moufang-Str.~1, 60438 Frankfurt am Main}

\begin{abstract}
We analyze how recent computations of the shear viscosity $\eta$ in the core of superfluid neutron stars affect the r-mode instability window.  We first analyze the contribution of superfluid phonons to the viscosity, both in their hydrodynamical and ballistic regime. We also consider the recent computation of $\eta$ arising from the collisions of electrons with electrons and protons by Shternin and Yakovlev, and discuss how the interactions among superfluid phonons and electrons might contribute to the shear viscosity. For assessing the r-mode instability window we compare the shear viscosity due to phonons in the hydrodynamical regime with respect to the shear viscosity due to electron collisions. Only at high temperatures the superfluid phonon contribution to $\eta$ starts to dominate the process of r-mode damping. While our results for the instability window are preliminary, as other dissipative processes should be taken into account as well, they differ  from previous evaluations of the r-mode damping due to the shear viscosity in superfluid neutron stars.

%We analyze how recent computations of  the shear viscosity $\eta$ in the core of superfluid neutron stars affect the r-mode instability window.
%We take into account the contribution of superfluid phonons to the viscosity, both in their hydrodynamical and ballistic regime. Moreover, we also consider the recent computation of $\eta$  arising from the collisions of electrons with electrons and protons by   Shternin and Yakovlev.
%dominated by transverse plasmon exchange. 
%We also discuss how the interactions among superfluid phonons and electrons might contribute to the shear viscosity, but argue that these interactions will be not be relevant for the study of r-mode damping.
%At very low temperatures, $T < 10^8$ K, the electron contribution to  $\eta$ dominates, as then phonons are in a ballistic regime, and they  mainly interact  with the crust of the star. 
%At higher temperatures the superfluid phonon contribution to $\eta$ is purely hydrodynamic and starts to dominate the process of  r-mode damping.
%While our results for the instability window
%are preliminary, as other dissipative processes should be taken into account as well, they differ substantially from previous evaluations of the r-mode damping due to the shear viscosity in superfluid neutron stars.   
\end{abstract}

\date{\today}
\maketitle

\section{Introduction}

Neutron stars are some of the most curious stellar objects in the Universe. They typically have 
 a mass of approximately $1.4 {\rm M_{sol}}$ and a radius of about $ 10$ Km, and supranuclear densities are thus expected in their
 cores.
  Our present understanding of
 the nuclear interactions at finite density suggests that neutron stars are composed by  well-defined 
different layers \cite{Lattimer:2006xb,Heiselberg:1999mq}. Thus, they are believed to have
a crust of about $1$ Km, with an outer part made of a lattice of ions embedded
in a liquid of electrons and an inner region made of nuclei embedded in a liquid both of electrons and 
 $^1S_0$ superfluid neutrons.  In the interior of the star nuclei are melted and both neutrons and protons 
are expected to condense into BCS-like superfluids. However, the neutron interaction
in the $^1S_0$ state at supranuclear matter density is repulsive, but it is
still possible to form
Cooper pairs in the $^3P_2$ channel. The proton density is
much smaller than the neutron density, and thus one expects    proton pairing in 
the isotropic $^1S_0$ channel, at least in some part of the core of the star. It is still unclear the content of the inner core of the star,
as it could be formed either by hyperonic matter,  phases with pion or kaon condensates, deconfined quark matter in a color superconducting phase, 
or other exotic dense matter phases \cite{Weber:2004kj}.

One possible way to verify the internal structure of a neutron star is through the observation of
the spectra of the different stellar oscillations, in a similar way as  asteroseismology is used to reveal
the internal structure of the Sun or other stars. In a neutron star there are different types of hydrodynamical 
oscillation modes. Of particular relevance are the r-modes, which are toroidal modes which only occur in rotating stars, the Coriolis force
acting as their restoring force. R-modes are generically unstable in all rotating stars through their coupling to  gravitational radiation (GR) emission~\cite{Andersson:2000mf,Lindblom:2001}.
When dissipative phenomena damp these r-modes the star can rotate without losing  angular momentum to GR. If dissipative phenomena are not strong enough,  these  oscillations 
will grow exponentially and the star will keep slowing down until some dissipation or non-linear mechanism   can  damp the r-modes. As most of the known neutron stars are pulsars, and their
rotation frequencies are known with a lot of accuracy, it has been recognized that the study of r-modes could effectively be used to constrain the internal stellar structure.

R-mode oscillations have been studied extensively in the literature and various
damping mechanisms have been proposed~\cite{Andersson:2000mf}. When viscous damping of the r-mode is taken into account the star is stable at low frequencies, or
at very low or high temperatures, but there is typically an instability region at enough high frequencies~\cite{Lindblom:1998wf, Andersson:1998ze}. The specific details
of this window depend on the model considered for the star.
Since the spin rate of several neutron stars falls in this instability window, most of the studies of
the r-modes are focused in detecting particular damping mechanisms that may modify or eliminate  the instability window. 
In particular, it has been claimed that surface rubbing between the
core and the crust of the star~\cite{Bildsten:1999zn,Glampedakis:2006mn} results in   a viscous boundary layer  which damps r-mode oscillations 
for sufficiently small frequencies. Other damping mechanisms  have been studied in the literature, as for example, the effect of mutual friction in superfluid neutron star \cite{Lindblom:1999wi,Lee:2002fp,Haskell:2009fz}.
It has also been recognized that if the r-mode amplitude gets sufficiently large, its growth should be stopped by non-linear dynamical effects \cite{Lindblom:2000az,Bondarescu:2007jw,Arras:2002dw,Lin:2004wx,Alford:2011pi}.  

It is  our goal to review the form of the r-mode instability window for neutron stars with  superfluid cores. Our main motivation comes from the fact that 
in previous studies of r-modes the contribution to transport phenomena arising from superfluid phonons had been ignored, so it was impossible to assess
its relevance for the r-mode damping.  All previous studies
of  r-mode oscillations of superfluid neutron stars are based in the hydrodynamical equations of two fluid motions, one 
corresponding to the neutrons, the other one to the charge-neutral fluid, made up by protons locked to electrons.
When finite temperature ($T$) effects are taken into account  one additional fluid motion has to be considered.
It is well-known that both fermionic and bosonic superfluid systems, 
like $^3$He or $^4$He, have additional degrees
of freedom, and that a two-fluid description of the superfluid is necessary to explain phenomena such as the second sound associated to thermal waves.
At low $T$ this second fluid gives account of the dynamics of some collective modes, the  superfluid phonons. The superfluid phonon is a Goldstone mode
which appears due to the fact that the  neutron condensate, regardless whether $^1S_0$ or $^3P_2$ neutron pairing is considered, spontaneously breaks the
 $U(1)$ baryonic symmetry. 
 The contribution of superfluid phonons to different transport coefficients has been computed for compact stars made of superfluid quark matter in the  color-flavor locked (CFL)  phase \cite{Alford:2007xm,Manuel:2004iv,Mannarelli:2009ia,Mannarelli:2008je}, together with their effect on the damping of r-modes \cite{Andersson:2010sh}. However, a similar analysis has not been yet carried out
for superfluid neutron stars.

 A study of  the collective modes in superfluid fermionic systems has shown
that the superfluid phonons may be overdamped, and even disappear
for some values of the temperature \cite{Leggett:1966zz, Leinson:2010ru}. Leggett found the conditions for the existence of phonons at finite $T$,
 relating their overdamping  to the values of the Landau parameters of the Fermi liquid. Based on 
recent calculations of the Landau parameters of neutron matter \cite{Benhar:2012jk,Holt:2012yv}, Leggett's criteria 
seem to indicate that phonons could be overdamped
at small values of the density of the star for some intermediate values of $T$, but exist 
at very low $T$ or close to the critical temperature for the superfluid transition. They would
exist for all $T$ 
in the interior of the core of
the neutron star, and  thus, contribute to the transport properties for those densities.
 A further study on the Landau parameters on $\beta$-stable nuclear matter is needed.
 In this work we will ignore the possibility of the disappearance of the superfluid phonons, as according to the
known values of the Landau parameters found in Refs. \cite{Benhar:2012jk,Holt:2012yv}, this could only happen in a regime where transport properties of the star
are dominated by the electrons.

%\%blue{The existence of the superfluid phonons and its damping at finite temperature was studied in superfluid Fermi systems \cite{Leggett:1966zz, Leinson:2010ru}. It was found that depending on the value
%of the Landau parameters of the Fermi liquid, these collective modes could be overdamped and even disappear for some intermediate values of the temperature.
%Very recent calculations on Landau parameters in neutron matter \cite{Benhar:2012jk,Holt:2012yv}  seem to indicate that phonons could be damped at  small values of the density of the star
%and  at some intermediate values of $T$,
% but exist in the interior of the core of the neutron star, and 
%and, thus, contribute to the transport properties. A further study on this existence criteria is needed.  In this work we will ignore the possibility of the disappearance of the superfluid phonons, as
%according to the known values of the Landau parameters quoted above, this could only happen in a regime where transport properties are dominated by the electrons. } 

In this article we consider a simplified model of the neutron star made up by neutrons, protons and electrons, using 
a  causal parametrization of the Akmal,  Pandharipande and  Ravenhall \cite{ak-pan-rav} (APR for short) equation of state (EoS) \cite{Heiselberg:1999mq} to describe the $\beta$-stable nuclear matter  inside the star. This EoS is a common benchmark for a nucleonic equation of state, which is widely used in neutron star matter calculations. 
In the core of the star, neutrons pair first in a $^1S_0$ channel, up to values of the particle density
$n \sim n_0 = 0.16$ fm$^{-3}$, the normal saturation density \cite{Lombardo:2000ec}. From this value, we consider that neutrons pair in the $^3P_2$ channel, all the way up to the center of the star. Thus,  in this work we don't consider the possibility of
exotic forms of dense matter in the inner core of the star.
 We assume a critical temperature for the transition from neutron superfluid to the normal phase in the range of $T_c \sim 10^{10}$ K.
% whereas we don't need the precise numerical value of the neutron gap.

%On the other hand, protons pair in the $^1S_0$ channel, up to critical densities of $n_c \sim (2.3-2.6) \ n_0$ \cite{Heiselberg:1999mq}. Thus, electromagnetic superconductivity only prevails up to that critical value of the particle density. Values of the $^1S_0$ proton gap are model dependent \cite{amundsen,wambach,baldo,chen,elgaroynpa96, elgaroyprl76,elgaroyprl77,Zuo:2004mc,Baldo:2007jx}. The result of Ref.~\cite{elgaroyprl77} suggests that this pairing survives up to $n \sim 2.3 \  n_0$ with a maximum value of $\sim 1 \,{\rm MeV}$ for $n \sim 1.3 \ n_0$. For definiteness, we use the model of  proton pairing of Ref.~\cite{elgaroyprl77}
 %in our computations. 

In this paper we make a first approach to evaluate whether the viscous damping associated to superfluid phonons in the hydrodynamical regime may affect the r-mode instability window of superfluid neutron stars, following our recent computation of Ref.~\cite{Manuel:2011ed}. We leave for the future  the study of how other dissipative effects associated to the superfluid phonons, such as including their contribution to the bulk viscosities coefficients  \cite{Manuel:2013bwa} or their impact on  mutual friction, might also modify  the r-mode instability window. In analyzing the r-mode instability window we also take into account the results for the electron contribution to the shear viscosity of Shternin and Yakovlev in Ref.~\cite{Shternin:2008es}, where it has been claimed that previous results of $\eta_e$ had overestimated its value. With all the above considerations we see that the r-mode instability window as arising only from shear viscous damping is modified 
from previous estimates. For a comparison with previous results with the same sort of simplified model for superfluid neutron stars, although with the use of a different EoS, see for example Ref.~\cite{Haskell:2009fz}.

This paper is structured as follows. In Sec.~\ref{shear-sec} we review  different computations of the shear viscosity in the superfluid core of neutron stars, first as arising from electron collisions with electrons or protons (Sec.~\ref{sec-shear-ee}), and then due to the phonon interactions with the boundary and among themselves  (Sec.~\ref{sec-shear-ph}).  While the computation of  $\eta$ of Ref.~\cite{Manuel:2011ed} was done under the assumption of neutron pairing in a $^1S_0$ channel,
we argue here that those results can as well be used for neutron pairing in a $^3 P_2$ channel, and explain how to incorporate the ballistic regime of the phonons in an effective shear viscosity coefficient. In Sec.~\ref{ph-e-sec} we discuss on how the electron-superfluid phonon interactions might contribute to the  shear viscosity, although we estimate that these contributions should not be the dominant ones for the r-mode damping. We analyze in Sec.~\ref{sec-rmode} how the shear viscosity associated to phonons in the hydrodynamical regime in Sec.~\ref{shear-sec} affect the r-mode instability window, modifying it with respect to previous results. We present a summary of our results and our conclusions in Sec.~\ref{conclu}. In Appendix \ref{appendixa} we show the computation of the coupling constant of the electron-superfluid phonon interaction in the low density limit.

We use natural units $\hbar = c = k_B=1$ in all intermediate equations unless otherwise stated. However we present numerical
values of the shear viscosity,  temperatures and lengths  in C.G.S. units.

%%%%%%%%%%%%%%%%%%%%%%%%%%%%%%%%%%%%%%%%%%%%%%%%%%%%%%
\section{Shear viscosity in the core of superfluid neutron stars}
\label{shear-sec}
%%%%%%%%%%%%%%%%%%%%%%%%%%%%%%%%%%%%%%%%%%%%%%

Knowledge of transport coefficients in the core of neutron stars is necessary in order to determine the possible damping mechanisms of different oscillation modes of the star.
In this Section we review the computations of the shear viscosity in the core of superfluid neutron stars, as due to electron collisions with electrons or protons (Sec.~\ref{sec-shear-ee}), and as due to phonons interacting among themselves or with the boundary, see Sec.~\ref{sec-shear-ph}.
We only consider those models that assume that  the electron contribution to the shear viscosity in the normal phase is higher than the nucleon contribution.
However, we note that there are some recent computations  \cite{Benhar:2009nr,Zhang:2010jf} that indicate that the nucleon contribution can be substantially increased due to many-body effects and three-nucleon forces and might overcome the electron contribution to the shear viscosity. We ignore such a possibility. As the electron collisions dominate over the nucleon collisions in the normal phase, one then expects that the same situation
occurs in the superfluid phase of the nuclear matter found in the star. For this reason, we don't give here the values of the contribution to the shear viscosity of the fermionic Bogoliubov quasiparticles, which, 
on the other hand, have not been computed so far.  In Sec.~\ref{ph-e-sec} we  study the superfluid phonon-electron interactions, and how these interactions
might contribute to the shear viscosity in the system, although we argue that they might be irrelevant for the study of the  r-mode damping. 
The values of the shear viscosity for phonons in the hydrodynamical regime presented in this Section  will be used in Sec.~\ref{sec-rmode} to assess  how they affect the r-mode instability window. We do not consider the contribution to the shear viscosity of phonon-Bogoliubov quasiparticle interactions, as we assume that phonon-phonon interactions dominate over phonon-quasiparticle processes because phonons are gapless modes.
It would be interesting to evaluate the shear viscosity coefficient due to the interaction of phonons with Bogoliubov quasiparticles, and check
whether this contribution has any impact on the r-mode instability window.

\subsection{Shear viscosity due to electron collisions}
\label{sec-shear-ee}

Most of the transport coefficients needed for the study of  the stellar evolution and structure of neutron stars have been originally studied by Flowers and Itoh in Ref.~\cite{Flowers}.
These authors evaluated the transport coefficients in the core of neutron stars, both assuming normal and superfluid phases. In the last case they noted
that the shear viscosity was mainly due to
electron collisions with electrons or protons. The pure nucleon contributions were, in turn, drastically suppressed. In their computation Flowers and Itoh assumed that the collisions
were dominated by
low momentum transfers, and made the approximation of assuming a dominant static  plasmon exchange. According to these authors
the value of the viscosity can be expressed in the extreme relativistic limit
as~\cite{Cutler} 
\be
\eta_e = \frac 13 \left(\frac{2}{5 \pi^2 \alpha T} \right)^2 p_e^5 \left [   \frac{\alpha}{\pi} \left(1 + \frac{ p_p m}{p_e^2 } \right) \right ] ^{1/2}  \ ,
\ee
where $p_e,p_p$ are the Fermi momentum of the electrons and protons, respectively, $m$ is the proton rest mass, and $\alpha$ is the electromagnetic fine structure constant.
Using the typical values of the Fermi momentum of the electrons and protons in the core of the neutron star, Cutler and Lindblom
realized that this expression could be well approximated by the simple formula
\be
\eta_e = 6.0 \times 10^6 \rho^2 \, T^{-2} {\rm \frac{g}{cm \cdot s}} \ ,
\label{ee}
\ee
where $\rho$ is the (local) mass density in ${\rm g/cm^3}$ and  $T$ stands for  the temperature  in Kelvin. This is the formula that has been typically used to estimate the r-mode instability window in neutron stars
with superfluid  cores.

Recently  Shternin and Yakovlev \cite{Shternin:2008es}  have reviewed all the computations of the electron shear viscosity in neutron star cores, both in  superfluid
and normal phases. These authors realized that in the computations of transport coefficients performed by Flowers and Itoh the charged particle collisions due to the exchange of transverse plasmons were erroneously screened with the same static dielectric function as the longitudinal interactions. 
%the role of the charged particle collisions due to the exchange of transverse plasmons had been neglected.
The transverse plasmon contribution enhances  the collisional rates with respect to the computation where only static longitudinal plasmons are considered, and this leads  to a reduction of the shear viscosity.
This fact had been first recognized by Heiselberg and Pethick, in the study of the transport coefficients for a degenerate quark plasma \cite{Heiselberg:1993cr}.
In the absence of neutron superfluidity and proton superconductivity, the electron collisions with electrons or protons give the leading contribution to $\eta$, and  this can be expressed as \cite{Shternin:2008es}
\be
\eta_{e}=\frac{n_e p_{e}}{5} \frac{1}{\nu_{e}}, \hspace{1cm} \nu_e=\nu_{ee}+\nu_{ep}+\nu'_{ee} \ ,
\label{etaee}
\ee
being $n_e$ the electron particle density and where we have used that  $p_{e}$ is equal to the electron chemical potential $\mu_e$ and to the electron effective mass $m_e^*$.  The electron effective collision frequency  $\nu_{e}$ results from the different collisions frequencies coming from the interaction of electrons with themselves ($\nu_{ee}$, $\nu'_{ee} $) as well as  with protons ($\nu_{ep}$) . Due to the different screening of the transverse and longitudinal contributions to the electron shear viscosity in the plasma of neutron stars, one should separate both contributions and this leads to the expressions  for the transverse $\nu^{\perp}_{e}$, longitudinal $\nu^{\parallel}_{e}$ and $\nu'_{ee}$ contributions of Ref.~\cite{Shternin:2008es}. For the transverse $\nu^{\perp}_{e}$, it is found that
\ba
\nu^{\perp}_{e}=\frac{\xi \pi \alpha}{4} \frac{q_t^{4/3}}{\pfe^2} T^{5/3} \ ,
\label{trans}
\ea
with $\xi=2 \Gamma(8/3) \zeta(5/3) (4/\pi)^{1/3} \approx 6.93$, being $\Gamma(x)$ the gamma function and $\zeta(x)$ the Riemann zeta function. The quantity $q_t$ is the characteristic plasma transverse wave number,  $q_t=\frac{2\sqrt{\alpha}}{ \sqrt{\pi}}\sqrt{\pfe^2+\pfp^2}$. For the longitudinal $\nu^{\parallel}_{e}$, one gets 
\ba
\nu^{\parallel}_{e}=\frac{4 \alpha^2}{\pfe^6} T^2 \ (I^{\parallel}_{e,e}+I^{\parallel}_{e,p}) \ ,
\label{long}
\ea
with
\ba
I_{e,e}^{\parallel}=&&\frac{\pi \pfe^6}{q_l} \  \ I_{2,e}(q_m/q_l) 
- \frac{ 3 \pi  \pfe^4  q_l}{4} \  I_{4,e}(q_m/q_l)  +\frac{3 \pi \pfe^2 q_l^3}{16}  \ I_{6,e}(q_m/q_l) 
    -\frac{\pi q_l^5}{64} \  I_{8,e}(q_m/q_l)  \ , \nonumber \\ 
 I_{e,p}^{\parallel}=&&\frac{\pi  \pfe^4  m^{*2}}{q_l}  \ I_{2,p}(q_m/q_l)  
- \frac{\pi (2 \pfe^2  m^{*2} + \pfe^4) q_l}{4} \  I_{4,p}(q_m/q_l) \nonumber \\
 &&+\frac{\pi (2 \pfe^2+m^{*2}) q_l^3}{16} \  I_{6,p}(q_m/q_l)  -\frac{\pi q_l^5}{64} \  I_{8,p}(q_m/q_l) \ , 
\ea
where $I_{e,e}(q_m/q_l)$ and $I_{e,p}(q_m/q_l)$ are
\ba
I_{e,(e,p)}(x)=\int^x_0 \frac{y^k}{(y^2+1)^2} dy ,
\ea
with $q_m=2 \pfe$, and  being $q_l=\frac{2\sqrt{\alpha}}{\sqrt{\pi}} \sqrt{m^* \pfp+\pfe^2}$ the Thomas-Fermi screening wave number. The effective mass of the proton $m^*$ is taken $m^*=0.8 m_n$, being $m_n$ the mass of the neutron \cite{Shternin:2008es}. Finally $\nu'_{ee}$ is given by
\ba
\nu'_{ee}= \frac{4 \alpha^2}{\pfe^6} T^2 I^{\perp \parallel}_{e,e} \ ,
\label{mix}
\ea
with
\ba
I_{e,e}^{\perp \parallel}= \frac{\pi  \pfe^2}{q_l}  \ \left[ \ \pfe^4 (I_{0,e}(q_m/q_l)+I_{2,e}(q_m/q_l)) 
 -\frac{ \pfe^2 q_l^2}{2} \ (I_{2,e}(q_m/q_l)+I_{4,e}(q_m/q_l))  + \frac{q_l^4}{16} \  (I_{4,e}(q_m/q_l)+ I_{6,e}(q_m/q_l)) \right]  \ .
\ea     
%and according to the authors of Ref.~\cite{Alford:2010fd}, 
%as
%\be
%\eta_e = 4.26 \times 10^{-26} (x_p n)^{14/9} T^{-5/3} {\rm \frac{g}{cm \cdot s}}  \ ,
%\label{eemod}
%\ee
%where $n$ is the baryon particle density in ${\rm cm^{-3}}$,  $x_p$ is the proton fraction and the temperature $T$ in Kelvin. 

Recently, this corrected value of the shear viscosity in the form given in Ref.~\cite{Alford:2010fd} has been used to analyzed the r-mode instability in neutron stars
with non-superfluid, non-superconducting cores ~\cite{Alford:2010fd,Vidana:2012ex}, as it was realized that also the electron contribution to $\eta$ dominates over the baryonic one
 %\footnote{\red{There are some recent calculations  \cite{Benhar:2009nr,Zhang:2010jf} that indicate that the nucleon contribution can be substantially increased due to many-body effects and three-nucleon forces and might overcome the electron contribution to the shear viscosity.}} 
 in
the normal phase of nuclear matter \cite{Shternin:2008es}.

Although the presence of neutron pairing does not affect directly the value of $\eta_e$, proton pairing does. This is so as protons also contribute to the photon polarization tensors.
Shternin and Yakovlev \cite{Shternin:2008es} found out that, on one hand, the proton superconductivity affects the plasma dielectric function and, hence, the screening of electromagnetic interactions whereas, on the other hand, implies an additional reduction of the electron-proton collision frequencies. Thus, the transverse collision frequency of Eq.~(\ref{trans}) is suppressed in proton superconducting matter  \cite{Shternin:2008es} as
\be
{\nu^{\perp}_{e}}^{s}=\nu^{\perp}_{e} R^{\perp}_{T} \ ,
\label{transsup}
\ee
with
\ba
&&R^{\perp}_{T}=\frac{1-g_1}{(1+g_3 \, y^3)^{1/9}}+
     (g_1+g_2) \, {\rm exp} [0.145-\sqrt{0.145^2+y^2}] \ , \nonumber \\
%\ea
%The different $g_1$, $g_2$ and $g_3$ constants read
%\ba
&&g_1=0.556 \ , \nonumber \\
&&g_2=0.426 \, y^{1/3}  +   0.0146 \, y^2 -  0.598 \, y^{1/3} \, {\rm exp}(-y) \ , \nonumber \\
&&g_3=16064 \, (1-g_1)^9 \ ,
\ea
The quantity $y$ is defined as $y=\frac{\Delta}{T}=\sqrt{1-\tau}\left(1.456-\frac{0.157}{\sqrt{\tau}}+\frac{1.764}{\tau}\right)$, where $\tau=T/T_{cp}$.  The variable $T_{cp}$ stands for the transition temperature from proton superconductor to normal fluid and is taken as $T_{cp}=10^9$ K \cite{levenfish}.
Moreover, the longitudinal dielectric function is almost insensitive to the presence of superconductivity and only the electron-proton collision frequency is suppressed  \cite{Shternin:2008es}. Then, Eq.~(\ref{long}) is modified according to
\ba
{\nu^{\parallel}_{e}}^s=\frac{4 \alpha^2}{\pfe^6} T^2 \ (I^{\parallel}_{e,e}+{I^{\parallel}_{e,p}}^s) \ ,
\label{longsup}
\ea
with
\ba
{I_{e,p}^{\parallel}}^s=R_p^{\parallel} \ I_{e,p}^{\parallel} \ ,
\ea
being the suppression factor $R_p^{\parallel}$ equal to
\ba
R_p^{\parallel}=&&[A^{\parallel}+(1.25-A^{\parallel}) \ {\rm exp}(-0.0437y)
      + (1.473 \ y^2+0.00618 \ y^4) \
     {\rm exp}(0.42-\sqrt{0.42^2+y^2})] \nonumber \\
&&  \times \  {\rm exp}(-\sqrt{0.22^2+y^2}) \ ,
\ea
and  $A^{\parallel}=1.45425$.  The  $\nu'_{ee}$ collision frequency is almost unchanged in proton superconducting matter \cite{Shternin:2008es}.

%dominated by the collisions mediated by transverse plasmons, which lead to the value
%\be
%\eta_e = \frac{27 \sqrt{3}  \pi^{1/3}}{40 \ T^2} \frac{n_e^2}{\alpha^{2/3} p_e} \left( \frac{\Delta}{p_p} \right)^{1/3}  \ ,
%\label{ee-proton}
%\ee
%where $n_e$ is the electron particle density and  $\Delta$ is the value of the $^1S_0$ gap for the protons.

In this work, in order to obtain the mass density, electron density, and proton and electron Fermi momenta, we use the APR EoS as common benchmark in a causal parametrization with parameters $\delta=0.2$ and $\gamma=0.6$, as seen in Ref.~\cite{Heiselberg:1999mq}.  

 %With regard to the $^1S_0$ gap for protons, results are model dependent \cite{amundsen,wambach,baldo,chen,elgaroynpa96, elgaroyprl76,elgaroyprl77,Zuo:2004mc,Baldo:2007jx}. The works \cite{elgaroynpa96, elgaroyprl76,elgaroyprl77} suggest a maximum value of approximately 1 MeV for densities of $\sim 1.3 \ n_0$ while it survives up to $n \sim 2.3 \ n_0$ \cite{elgaroyprl77}, although the inclusion of three-body forces, self-energy effects and the polarization interaction might reduce the domain of pairing \cite{Baldo:2007jx}.  In this paper we use the estimate of Ref.~\cite{elgaroyprl77}. 

\begin{figure}[t]
\begin{center}
\includegraphics[width=0.7\textwidth]{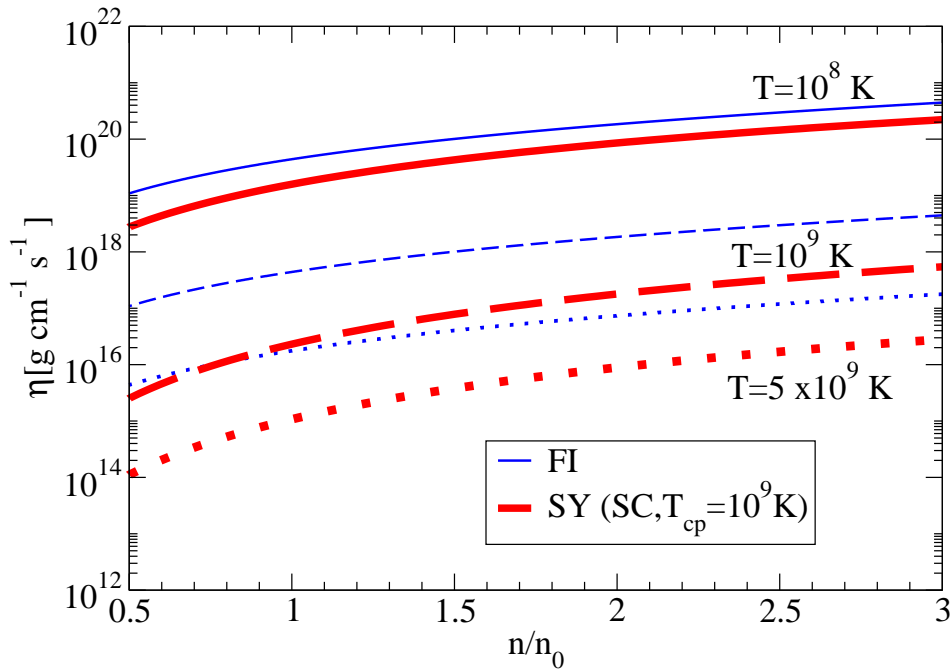}
\end{center}
\caption{(Color online) Shear viscosity due to electron collisions assuming static photon screening from Flowers and Itoh (FI)
[Eq.~(\ref{ee})] and due to the inclusion not only of the longitudinal plasmon exchange but also taking into account the transverse plasmon one from Shternin and Yakovlev (SY) in proton superconducting matter (SC) with a transition temperature for protons of $T_{cp}=10^9$ K [Eqs.~(\ref{etaee},\ref{mix},\ref{transsup},\ref{longsup})]. The plots are given as a function of the particle density  in units of normal saturation density, $n_0=0.16 \ {\rm fm^{-3}}$, for three different temperatures, $T=10^8 {\rm K}$ (solid lines), $T=10^9 {\rm K}$ (dashed lines) and $T=5 \times 10^{9} {\rm K}$ (dotted lines).}
\label{fig:ee}
\end{figure}

In Fig.~\ref{fig:ee} we show the contribution to the shear viscosity given by electron scattering with electrons and protons (in the following we will refer them solely as electron collisions) as a function of the particle density in units of normal saturation density, $n_0=0.16 \ {\rm fm^{-3}}$, for three different temperatures, $T=10^8 {\rm K}$ (solid lines), $T=10^9 {\rm K}$ (dashed lines) and $T=5 \times 10^{9} {\rm K}$ (dotted lines).  A particle density of  $n_0=0.16 \ {\rm fm^{-3}}$ corresponds to a mass density of $\rho_0 \sim 2.8 \times 10^{14} \ {\rm g/cm^3}$. A temperature of $T \sim 5 \times 10^{9} {\rm K}$ is close to the expected transition temperature to normal fluid in case of neutron superfluidity  ($T_c \sim 0.57 \ \Delta$).
%in case of proton superconductivity ($T_c \sim 0.57 \ \Delta$).
We present the predictions of $\eta_e$ by Flowers and Itoh (FI), which assumed static photon screening,
and which are expressed in  Eq.~(\ref{ee}), and also the predictions of Shternin and Yakovlev (SY) in proton superconducting matter that consider not only the longitudinal but also the transverse contributions to the electron shear viscosity  in Eqs.~(\ref{etaee},\ref{mix},\ref{transsup},\ref{longsup}). We observe that, in accordance with the statements of Shternin and Yakovlev, the old values of
 the shear viscosity used in almost all studies of r-modes have been overestimated by a few orders of magnitude. 
 
%which assumed that $\eta_e$ is dominated by transverse plasmon exchange.
% In the last case, we assume that proton pairing only prevails for densities in the range  $n < n_c \sim 2.3 \ n_0$. Thus, we use the values of Eq.~(\ref{ee-proton}) up to that critical value, and use Eq.~(\ref{eemod}) for higher densities, where no proton pairing occurs. Note that there is a clear discontinuity in $\eta_e$ in this case, but this is an artifact arising from the fact that Eq.~(\ref{ee-proton}) is not valid in the regime where the proton gap
%vanishes, $\Delta \rightarrow 0$. The two curves should interpolate continuously, however the  formula for $\eta_e$ in the case of almost  vanishing $\Delta$ is not known. Because the discontinuity only affects the value of $\eta$ for the critical value of the density, then it won't affect much our analysis of the impact of $\eta$ on
%the r-mode instability window, where one integrates the values of $\eta$ for all the densities in the star.  

% We also observe that proton pairing leads to a decrease of the shear viscosity,  as one could naturally expect.

\subsection{Shear viscosity due to  phonons in superfluid neutrons stars including finite size effects }
\label{sec-shear-ph}

The phonon contribution to the shear viscosity, assuming neutron pairing in a  $^1S_0$ channel, has been recently computed in Ref.~\cite{Manuel:2011ed}.
We review here  the results of that reference, and further argue that they can be  extended
for phases with $^3P_2$ neutron pairing, where superfluid phonons also exist. We also comment how finite size corrections have to be taken into account when  the phonon mean free path becomes larger than the size of the core of the star, and thus phonons are
in a ballistic, rather than hydrodynamical, regime.

The basic phonon interaction rates can be derived using effective field theory techniques. The effective phonon Lagrangian is  presented as an expansion in derivatives (or momenta and frequencies) of the phonon field, the terms of this expansion being restricted by symmetry considerations. The coefficients of the Lagrangian can be  computed from the microscopic theory, through a standard
matching procedure, and thus  they depend on the short range physics of the system under
consideration. The effective field theory assumes that the phonon momentum $q$ and frequency $\omega$ are such that $q v_F, \omega  \ll \Delta$, where $v_F$ is the Fermi velocity.
The expansion parameter of the theory is not a coupling constant, bur rather the values of $q v_F/ \Delta$ and $\omega / \Delta$.

The leading order  phonon Lagrangian in a derivative expansion turns out to be universal for all those superfluids sharing the same global symmetries. It
 reads~\cite{Son:2005rv}
\begin{equation}
\label{LO-Lagran}
\mathcal{L}^{\rm LO}_{\rm ph} =P (X) \ , \qquad  X = \mu_n-\partial_t\varphi-\frac{({\bf \nabla}\varphi)^2}{2m} \ ,
\end{equation}
where $P(\mu_n)$ and $\mu_n$ are the pressure and neutron chemical potential, respectively, of the superfluid at $T=0$, and
 $\varphi$ is the phonon field.
We thus see that the leading phonon interactions depend on the EoS of the star. 
The phonon contribution to the shear viscosity $\eta$ is obtained by considering small departures from equilibrium to the phonon distribution function and linearizing the corresponding transport equation \cite{Manuel:2011ed}. It is then necessary to use variational methods in order to solve the transport equation, as in Refs.~\cite{Manuel:2004iv,Alford:2009jm,Rupak:2007vp}. The final expression for the shear viscosity is given by \cite{Manuel:2011ed}
\be
\eta=\left( \frac{2 \pi}{15} \right)^4 \frac{T^8}{c_s^8} \frac{1}{M} \ ,
\ee  
where $c_s$ is the speed of sound and $M$ represents a multidimensional integral that contains the thermally weighted scattering matrix for phonons. It turns out that the shear viscosity is dominated by binary collisions, and it was found that, as it happens in other superfluid systems, $\eta \propto 1/T^5$, the proportionality factor depending on the EoS of the star.
%The phonon contribution to the shear viscosity $\eta$ turns out to be dominated by binary  collisions, and it was found that, as it happens in other superfluid systems, $\eta \propto 1/T^5$, the proportionality factor depending on the EoS of the star.
 For the particular computations of Ref.~\cite{Manuel:2011ed}, the
 EoS for $\beta$-stable nuclear matter  obtained by APR \cite{ak-pan-rav}, in a causal parametrization introduced by  Heiselberg and Hjorth-Jensen \cite{Heiselberg:1999mq}, was used.  The APR EoS ignores  the effect of neutron pairing. However, because neutron pairing only affects
those neutrons which are closed to their Fermi surface, one does not expect that the effect of pairing might have a big impact in the EoS. This effect should be of the order $\frac {\Delta^2}{\mu_n^2}$, where 
$\Delta$ is the value of the s-wave gap. Given the fact that the neutron gap is believed to reach values $\Delta \sim 3$ MeV (although correlations tend to reduce its value \cite{Lombardo:2000ec}) while $\mu_n \sim 900$ MeV,  this seems to be a good approximation.

It is possible to construct using the same techniques a  next-to-leading order (NLO) phonon Lagrangian ~\cite{Son:2005rv}. One of the most remarkable changes of the phonon physics at NLO is that the phonon dispersion law is then modified. We have evaluated, through a matching procedure with the underlying baryonic theory, that the NLO correction to the phonon dispersion law  is  $q^2 v_F^2/(45 \Delta^2)$ \cite{sarkar,Manuel:2013bwa}. For a BCS system, we can write as an estimate that the critical temperature is $T_c \sim 1/2 \Delta$, with $\Delta$ being evaluated at zero temperature. Since in any computation for the transport coefficients the phonons contributing the most have momenta $q \sim T/c_s$, we are lead to conclude
that contributions of the NLO pieces can be neglected when $T <<  2 \sqrt{45} c_s/ (v_F T_c)$. For the explicit values of $c_s$ and estimates for $v_F$, we find that the NLO corrections are very small even close at $T_c$. We thus find that the LO phonon Lagrangian gives a very good estimate of the phonon contribution to $\eta$ at temperatures even close to the temperature
of the superfluid phase transition $T_c$. For this reason we will take the value of the viscosity computed with the LO Lagrangian as a good approximation for high values of $T$.

We now argue that the results of Ref.~\cite{Manuel:2011ed} for the contribution of the superfluid phonons to $\eta$ are as well valid for the anisotropic $^3 P_2$ neutron pairing, if one also ignores corrections of order $\frac {\bar \Delta^2}{\mu_n^2}$, where 
${\bar \Delta}$ is the angular average of the $^3 P_2$  gap function.  At this order in the approximation the breaking of rotational invariance due to the existence of an anisotropic condensate is not felt by the superfluid phonons. This fact has been explicitly checked in Ref.~\cite{Bedaque:2003wj} for $^3 P_2$ neutron pairing, after computing the superfluid phonon kinetic term, integrating out
the neutron degrees of freedom, and realizing that at that order one gets the same results that in a rotational invariant (s-wave) superfluid. 
However, one should be aware of a relevant  difference between the s-wave and p-wave superfluid phases. In the last case there is also a spontaneous symmetry breaking of the rotational symmetry, and in this case there are additional  massless collective excitations. The Goldstone modes associated to the breaking of rotational invariance, the so-called angulons \footnote{The existence of the angulons is, however, under debate \cite{Leinson:2012pn}}, should as well contribute to the shear viscosity.  We leave that analysis for a future project. In fact, the anisotropy in the hydrodynamical equations describing the p-wave superfluid phases of the core of neutron stars has not been considered so far in the study of any of its possible oscillation modes. Our results for $\eta$ when there is  $^3P_2$ neutron pairing should be considered as a lower bound in this case.

The phonon contribution to  $\eta$ found out in Ref.~\cite{Manuel:2011ed}  diverges at sufficiently low temperatures. However, this is an unphysical result, which does not take into consideration that  finite size effects actually prevent the viscosity from increasing indefinitely at low temperature. In Ref.~\cite{Manuel:2011ed} it was actually shown that at sufficiently low $T$ the phonon mean free path typically exceeds the value of the radius of the star. Let us define the viscous mean free path by the formula
\be
\label{bulkvis}
\eta_{\rm bulk} = \frac 15 \rho_{\rm ph}  c_s  l_{\rm ph} \ ,
\ee
where $\eta_{\rm bulk}$ is the phonon contribution to the shear viscosity in the hydrodynamical regime and
 $\rho_{\rm ph} = \frac{2 \pi^2 T^4}{45 c_s^5}$ is the phonon contribution to the mass density. For hydrodynamics to be valid, one demands  the Knudsen number $K_n$ to be small.  The Knudsen number is defined as the ratio of the mean free path versus the typical macroscopic scale, in our case, the radius of the core of the star $R_c$, thus $K_n = \frac{l_{\rm ph}}{R_c}$. Following Fig.~5 of Ref.~\cite{Manuel:2011ed}, the Knudsen number is $K_n \lesssim 1$ for values of $T \gtrsim 5 \times 10^8-10^9$ K (the precise threshold $T$ depends
strongly on the value of the nucleon particle density one considers). For $K_n > 1$ an hydrodynamic description of the phonons is impossible, and one should write a Boltzmann equation to describe the phonon dynamics.

When the phonon mean free path exceeds the size of the superfluid core of the star, the transport properties of the phonons are mainly governed by their interactions with the crust, which is the boundary of the superfluid region, rather than by their self-interactions. For simplicity, we will neglect the structure of the crust, and assume that the crust is not in a superfluid phase. This is a sort of approximation that has been typically done in all  studies of the r-modes oscillations, as even it is considered that r-mode damping might be dominated by the rubbing effect of the fluid with either a rigid or elastic crust \cite{Bildsten:1999zn,Glampedakis:2006mn}.  We  make here the same assumption in considering that the superfluid phonon motion is confined to the core of the star. A much more rigorous treatment should 
consider that the superfluid phonons of the core  can penetrate into the inner crust, where  neutrons also pair in a s-wave condensate. This would also imply to consider that a transfer of neutrons between the inner crust and outer core is indeed possible, which should have consequences in the r-mode evolution, through the so-called rocket effect \cite{Colucci:2010wy}. In this article we will take the simpler assumption by considering the crust as a rigid boundary.
If the superfluid phonon motion is confined to the core of the neutron star, then phonons reaching the crust  can be either  absorbed or they can be scattered back. If the phonons are diffused, they will exert a shear stress on the boundary, in the same way as the particles in a rarefied gas produce shear stress when diffused by the walls of their container. 

In the case described above,
when $K_n >1$,  phonons are in a ballistic regime, where no  hydrodynamical description of their behavior is possible. However, it has been observed experimentally that for $^4$He at very low temperature ballistic phonons can still efficiently damp the movement of immersed objects, see Refs.~\cite{Eselson,Niemetz,Zadorozhko}. This damping can still be described by an effective or ballistic 
shear viscosity coefficient defined by
\be\label{etaball-He4}
\eta_{\rm ball}  \equiv \frac{1}{5}  \rho_{\rm ph}\chi  c_s  d\,,
\ee
where $d$ is the typical size of the oscillating object, and $\chi$ is a coefficient that measures the probability  that the phonons are diffused at the boundary with the object. This expression leads to excellent agreement with the experimental data for large Knudsen numbers \cite{Zadorozhko}.

In the intermediate regime between large or small Knudsen numbers, that is, between the ballistic and hydrodynamic regime, one can effectively calculate the shear viscosity according to
\be
%\label{eff-eta}
\eta_{\rm eff} = \left( \eta_{\rm  bulk}^{-1} + \eta_{\rm ball}^{-1}\right)^{-1} \, .
\label{eff}
\ee
This formula can be justified from the fact that the shear viscosity is proportional to a collisional relaxation time,  and that if one considers different processes occurring independently with different relaxation times, the total relaxation frequency is the sum of the partial relaxation frequencies. The same sort of formula has been used recently to describe the low $T$ behavior of the experimental values of the shear viscosity in the superfluid phase of a cold Fermi atomic gas in the unitarity limit, where
one assumes that it is also dominated by the dynamics of the superfluid phonons \cite{Mannarelli:2012su, Mannarelli:2012eg}.

 We will use Eq.~(\ref{eff}) to give account of the phonon contribution to $\eta$ in the core of the star, taking for $d$ the value of the radius of the core. The radius of the core is fixed by the transition between the crust and the core, which corresponds to  $n \sim 0.5 \  n_0$.
% {\bf we should say which value of $R_c$ we use- maybe here how we find it, i suppose it will be different for the two stars considered in the paper; which value did you use  in fig.2?}.
 The value of $\chi$ could be obtained if at a microscopic level one could determine the sort of scattering processes that the phonons experience when colliding with the crust. Here we will assume that  $\chi \sim 1$.
We take for $\eta_{\rm  bulk}$ the values of the shear viscosity computed in Ref.~\cite{Manuel:2011ed}, which are strictly valid in the hydrodynamic, or low Knudsen number, regime.
Note that $\eta_{\rm eff}$ gets dominated by the smaller contribution among $\eta_{\rm bulk}$ and $\eta_{\rm ball}$, so it describes properly the different hydrodynamical, transition and  ballistic regimes.
\begin{figure}[t]
\begin{center}
\includegraphics[width=0.7\textwidth]{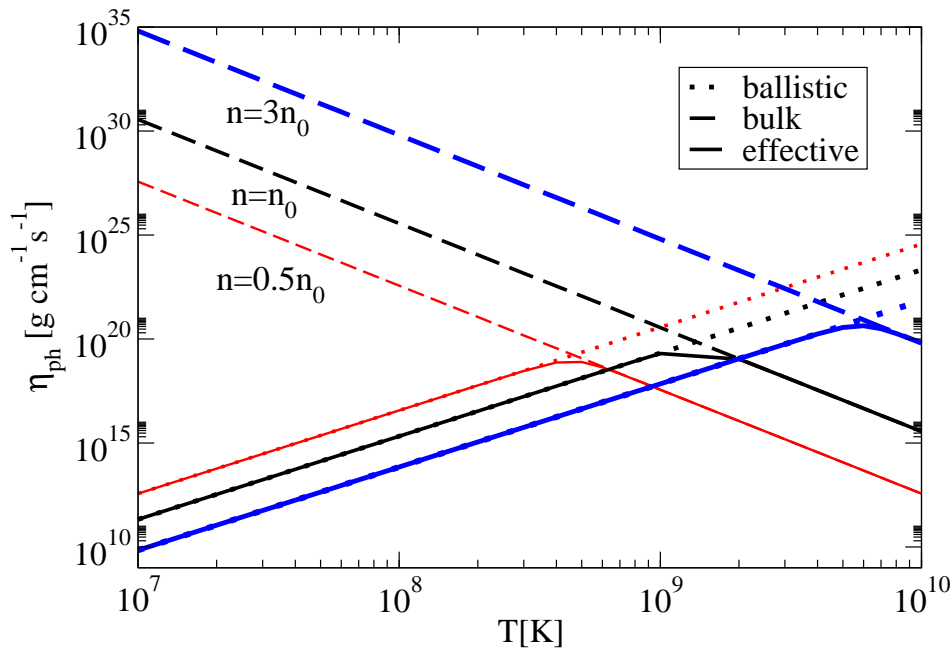}
\end{center}
\caption{(Color online) Phonon contributions to the shear viscosity as a function of temperature for different particle densities in units of normal saturation density, $n_0=0.16 \ {\rm fm^{-3}}$. We plot separately the ballistic viscosity 
$\eta_{\rm ball}$  (dotted lines), the hydrodynamical viscosity $\eta_{\rm bulk}$ (dashed lines) and the effective description $\eta_{\rm eff}$ (solid lines), which takes  into account properly  which regime dominates the dissipation.
 }
\label{fig:ph}
\end{figure}

In Fig.~\ref{fig:ph} we show values of the ballistic  shear viscosity  $\eta_{\rm ball}$ (dotted lines), of the hydrodynamic shear viscosity $\eta_{\rm bulk}$ (dashed lines) and the effective viscosity  $\eta_{\rm eff}$ (solid lines) that allow to describe all the regimes (ballistic, hydrodynamical, and the transition among the two). These contributions are displayed as a function of temperature for different particle densities in units of the normal saturation density. The densities studied are those expected to be reached in the core of neutron stars. The ballistic contribution in Fig.~\ref{fig:ph} is calculated for a core radius $R_c =9.96$ Km, which corresponds to a density of 0.5 $n_0$ for a star of 1.93 ${\rm M_{sol}}$. 

We observe three distinct regions depending on the dominant contribution to the shear viscosity.  For $T < 10^{8} {\rm K}$ the shear viscosity is described in terms of the ballistic contribution for all densities in the core. As the temperature increases, the transition from the ballistic to the hydrodynamic domain takes place. This intermediate regime is strongly dependent on the density, as we can see in Fig.~\ref{fig:ph}. For $n \sim n_0$ this transition domain appears for $T\sim 10^{9} {\rm K}$, while for higher temperatures, the hydrodynamic regime dominates.  As indicated before, the validity of the hydrodynamical description was already studied in Ref.~\cite{Manuel:2011ed} by means of the analysis of the phonon mean free path as compared to the radius of the neutron star. As one would expect, this study and that of Ref.~\cite{Manuel:2011ed} are in good agreement.
We also observe that for a fixed value of the temperature the ballistic viscosity decreases  as soon as the particle density increases, while the hydrodynamical viscosity exhibits the opposite behavior.

\begin{figure}[t]
\begin{center}
\includegraphics[width=0.7\textwidth]{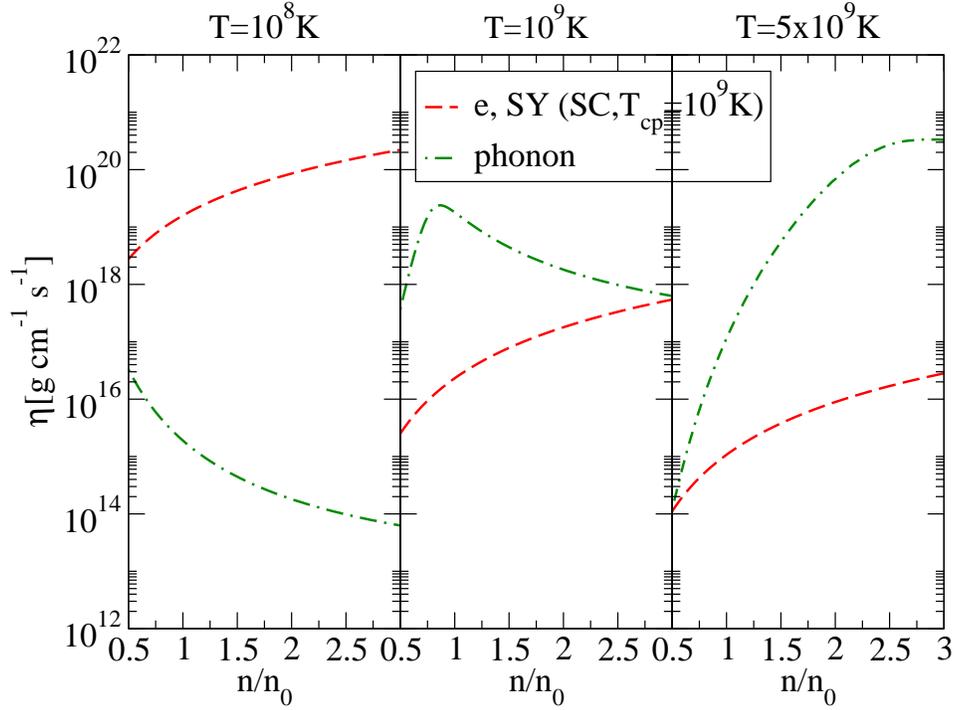}
\end{center}
\caption{(Color online) Shear viscosity resulting from electron collisions from Shternin and Yakovlev in proton superconducting matter [Eqs.~(\ref{etaee},\ref{mix},\ref{transsup},\ref{longsup})]  together with the effective description of the phonon shear viscosity [Eq.~(\ref{eff})], as a function of the particle density  in units of normal saturation density, $n_0=0.16 \ {\rm fm^{-3}}$, for three different temperatures, $T=10^8 {\rm K}$ (left panel), $T=10^9 {\rm K}$ (middle panel) and $T=5 \times 10^{9} {\rm K}$ (right panel).}
\label{fig:ee-ph}
\end{figure}

The question that remains is whether, in superfluid neutron stars, the phonon shear viscosity  could govern the r-mode damping inside the core. As discussed in Sec.~\ref{sec-shear-ee}, it is believed that the electron contribution to the shear viscosity dominates with respect to any other baryonic contribution. In Fig.~\ref{fig:ee-ph}, we present the predictions from Shternin and Yakovlev for the shear viscosity given by electron collisions in proton superconducting matter  [Eqs.~(\ref{etaee},\ref{mix},\ref{transsup},\ref{longsup})]  together with the effective phonon shear viscosity, where the ballistic and hydrodynamical regimes are both considered [Eq.~(\ref{eff})]. These are shown as  a function of the particle density for three different temperatures, $T=10^8 {\rm K}$  (left panel), $T=10^9 {\rm K}$ (middle panel) and $T=5 \times 10^{9} {\rm K}$ (right panel).

We concentrate now on the analysis of the phonon shear viscosity and the comparison with the viscosity coming from electron collisions. We observe that the effective description of the phonon shear viscosity strongly depends on the temperature. For $T \sim 10^8 {\rm K}$ (left panel of Fig.~\ref{fig:ee-ph}), the ballistic description dominates and the viscosity scales as $1/c_s^4$ with density. However, as the  temperature increases, the transition between the ballistic and hydrodynamic domains takes place. For $T\sim10^9 {\rm K}$ (middle panel of Fig.~\ref{fig:ee-ph}), we find that the phonon shear viscosity shows a strong density dependence coming from this transition.  As seen in Fig.~\ref{fig:ph}, low densities reach faster the hydrodynamic domain. Thus, for $T\sim10^9 {\rm K}$ and $n\lesssim n_0$,  the shear viscosity increases rapidly as function of the density following a hydrodynamical behavior. This  is followed by a decrease with density as the ballistic description takes over. It is interesting to note that for  $T\sim10^9 {\rm K}$ and $n \lesssim 3 n_0$, phonons interacting among themselves and with the boundary provide a higher dissipation  than electron collisions. For $T \sim 5 \times 10^{9} {\rm K}$  (right panel of Fig.~\ref{fig:ee-ph}), the description of the phonon shear viscosity is hydrodynamic and phonon-phonon scattering dominates over electron collisions in the whole core. Therefore, we conclude that processes involving phonons, in particular in the hydrodynamical regime, can have a dominant role in the dissipation in neutron stars for some values of the temperature, and affect, in turn, certain oscillation modes of the star, such as the r-modes. We will address this issue in Section~\ref{sec-rmode}.

%%%%%%%%%%%%%%%%%%%%%%%%%%%%%%%%%%%%%%%%%%%%%%%%%%%
\subsection{Interactions among superfluid phonons and electrons in the core of neutron stars}
\label{ph-e-sec}
%%%%%%%%%%%%%%%%%%%%%%%%%%%%%%%%%%%%%%%%%%%%%%%%%%% 
 
 \begin{figure}[t]
\begin{center}
\includegraphics[width=0.6\textwidth]{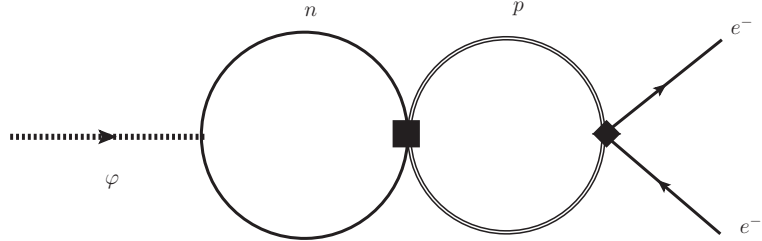}
\end{center}
\caption{Feynman diagram for electron-phonon scattering in matter with protons and neutrons forming $^1S_0$ Cooper pairs.}
\label{fig:diag}
\end{figure}

We study here the microscopic interactions of the superfluid phonons with the other light degrees of freedom of the core of the neutron star, the electrons. This may allow us to see whether these processes can give a substantial contribution to the shear viscosity.
%and ultimately, to clarify whether the phonon and electron fluid motions can equilibrate in short time scales.

Let us  assume that both neutrons and protons form Cooper pairs in a $^1S_0$ channel (a similar reasoning can be applied for the p-wave superfluid case).
Effective field theory techniques can be used to write down the Lagrangian describing the electron  and phonon interactions.  At leading order in a derivative expansion, simple symmetry arguments suggest that this is given by
\be
\label{Lag-phon-el}
{\cal L}_{\rm ph-el} = g \left (  \bar \psi_e \gamma^0 \psi_e \partial_0 \varphi + a_s \bar \psi_e \gamma^i \psi_e \partial_i \varphi\right) \ ,
\ee
where $\psi_e$ is the electron field.
The values of the coupling constants $g$ and $a_s$ can be obtained after integrating out  the neutron, the proton and the photon from the complete underlying microscopic field theory. In particular, their values can be obtained in the low density domain
after the evaluation of the Feynman diagram depicted in Fig.~\ref{fig:diag} (see Appendix \ref{appendixa} for details). At low density one finds
\be
\label{coupling-phel}
g  \approx  \frac{2 \,p_{n} m \,c_s}{\pi \sqrt{\rho_n}} a_{np} \ ,
\ee
where $p_n$ if the neutron Fermi momentum, $\rho_n$ is the neutron mass density, and
$a_{np}$ is the neutron-proton scattering length, that in vacuum  takes the value $a_{np} = -23.7$ fm. 
One might expect, however, that at high density the above value of $g$ might be substantially corrected.
We will assume that $a_s \sim 1$.

In the superfluid core of a neutron star the electron chemical potential  $\mu_e$ has values of a few hundreds of MeV, 
while the temperature is very low, $T \ll 1$ MeV. So it is a good approximation to consider that  electrons are totally degenerate, and neglect how thermal effects may modify their Fermi sea.
The interactions of the phonons with the electrons inside the Fermi sea are very suppressed, and a phonon, which typically carries an energy of order $T$ with $T \ll \mu_e$, can hardly excite an electron  inside the Fermi sea. 
However, we note that phonons can interact with the electrons which lie on their Fermi surface.

Let us see now how knowledge of Quantum Electrodynamics (QED) in the ultradegenerate limit can be used to infer properties of  the superfluid phonon-electron interactions.
For that purpose, it is enough to realize that the Lagrangian of Eq.~(\ref{Lag-phon-el}) can also be written as 
\be
\label{Lag-phon-el-QED}
{\cal L}_{\rm ph-el} = g  \bar \psi_e \gamma^\mu \psi_e {\tilde A}_\mu \ ,
\ee
if we define ${\tilde A}_\mu = ( \partial_0 \varphi,  a_s \partial_i \varphi)$. In this way, this interaction term  resembles  that of the gauge interaction of the electrons with an electromagnetic field potential.
The phonon self-energy can thus be obtained from the knowledge of the photon self-energy. In momentum space, with wave vector $k_\mu = (k_0, {\bf k})$, one has
\be
 \frac 12 {\tilde A}_\mu (-k) \Pi_{\rm QED}^{\mu \nu}(k)  {\tilde A}_\nu (k) = - \frac 12 \varphi (-k) {\tilde k}_\mu \Pi_{\rm QED}^{\mu \nu} (k) {\tilde k}_\nu \varphi (k) \ ,
\ee
where we have defined the wave vector $ {\tilde k}_\mu = (k_0, a_s {\bf k}) $, and we take the QED expressions for the polarization tensor simply replacing the electromagnetic coupling constant $e$ by $g$.
Expressions for the polarization of QED in the limit of high chemical potential $\mu_e$ and vanishing temperature are well-known
for values of the momenta much less than $\mu_e$. These are given by the so called hard dense loops (HDL), and it can be shown that only the electrons on their Fermi surface contribute to
the polarization tensor \cite{Manuel:1995td}.  We thus see that the superfluid phonon polarization tensor is expressed in terms of the photon polarization tensor according to
\be
\Pi (k) = - {\tilde k}_\mu   {\tilde k}_\nu \Pi_{\rm QED}^{\mu \nu} (k) \ .
\ee

The HDL polarization tensor contains an imaginary part which gives account of the so-called Landau damping \cite{Landau}. These are processes where a photon is absorbed (or emitted) by the electrons.
%This is a collisionless dissipation which does not involve an increase of entropy, and thus such a decay does not contribute directly in the evaluation of the transport coefficients of the system.
Based on the parallelism we have just traced of the photon and phonon interactions with the electrons, we are then lead to conclude that the phonons in the superfluid nuclear media also suffer
Landau damping.   The phonon damping rate, which is directly related to Landau damping, is given  by
\be
\gamma =  \frac{{\rm Im} \, \Pi(k_0= E_k, {\bf k})}{2 E_k} = g^2 c_s^2 (1+a_s)^2 \frac{\mu_e^2}{2 \pi^2} k \ ,
\ee
where we have used that the phonon dispersion law is $E_k = c_s k$ at leading order. 

Because Landau damping does not involve an increase of entropy, it cannot 
contribute directly to the evaluation of the shear viscosity. Following the analogy with the QED case, one realizes that 
electron collisions mediated by one-phonon exchange will turn out to be the most relevant process involving both electrons and phonons that might contribute
to the shear viscosity.
 We leave that computation for a future project, but present below some arguments of why our conclusions for the evolution of r-modes might not be affected by this process.
 %some reasoning of why not take them into account might not affect our conclusions for the evolution of r-modes. 
The computation of the shear viscosity as due to the collisions mediated by one-photon exchange or one-phonon exchange 
in the degenerate plasma turns out to be very similar. For the QED case the scattering matrix is given by \cite{Heiselberg:1993cr}
\be
{\cal M}  \sim e^2 J^e_\mu D^{\mu \nu}(k) J^{e'}_\nu \ ,
\ee
where $J^e_\mu$  and $J^{e'}_\nu$ are the electromagnetic currents associated to the two electrons participating in the collision, and $D^{\mu \nu}$ is the photon propagator.
In the electron-phonon case the scattering matrix reads
\be
{\cal M}  \sim g^2 J^e_\mu {\tilde k}^\mu {\tilde k}^\nu G(k) J^{e'}_\nu \ ,
\ee
where $G(k)$ is the phonon propagator.
Because the energy transfer in these collisions is at most of order $T$ \cite{Heiselberg:1993cr}, one is lead to conclude that the relevance of one process versus the other one
is provided by the ratio $e^2 / (g^2 T^2)$. 
Within the low-density approximation for the evaluation of the coupling constant $g$ of Eq.~(\ref{coupling-phel}), we find that this coupling is at most 0.1~MeV$^{-1}$. We conclude that at  temperatures $T \lesssim (10^8-10^9)$K,  the electron collisions mediated by one-photon exchange seem to be more relevant than those mediated by one-phonon exchange, at least in the low-density domain. Thus, we will neglect in this paper the shear viscosity  arising from the interactions among electrons and the superfluid phonons, although  a much more detailed discussion about the superfluid phonon-electron interactions will be presented somewhere else, in particular in the high-density region.

 It is also interesting to note that the phonon damping rate, which is given by the ratio of the Landau damping $\gamma$ with respect to the phonon dispersion law at LO,  turns out to be $\gamma/(c_s k) \lesssim$ 1 for densities $n \lesssim  0.12\  {\rm fm}^{-3}$. For higher densities beyond normal saturation density, the estimate for the phonon damping rate should be corrected. Among others, one should go beyond the low-density theorem for the nucleon-proton interaction and add higher-order density corrections. Thus, further analysis in the high-density domain is needed in order to determine whether phonons can be efficiently damped well inside the core.

%%%%%%%%%%%%%%%%%%%%%%%%%%%%%%%%%%%%%%%%%%%%%%%%%%%%%%%%%%%%%%
 \section{  Shear viscosity and  r-mode instability window}
 \label{sec-rmode}
%%%%%%%%%%%%%%%%%%%%%%%%%%%%%%%%%%%%%%%%%%%%%%%%%%%%%%%%%%%%%%

In order to assess the r-mode instability window one has to compute the value of the different time scales associated to  
 both the instability due to gravitational wave emission $\tau_{\rm GR}$ , and to the various dissipative processes that may damp the r-mode in the star.
 In a superfluid neutron star the critical rotation rate is obtained by studying the equation
\be
 \frac{1}{\tau (\Omega,T)} = - \frac{1} { |\tau_{\rm GR} (\Omega) |} +\frac{1}{\tau_{\eta} (T)} +\frac{1} {\tau_{\zeta} (T)} + \frac{1} {\tau_{\rm MF} (\Omega)} \ ,
 \ee
where
$\tau_{\eta} ,\tau_{\zeta}$ and $\tau_{\rm MF}$ refer to the damping times associated to processes that involve the shear viscosity, the bulk viscosity and
mutual friction, respectively. For superfluid neutron stars  the bulk viscosity damping time has been assumed to be negligible  in the literature. Only
the effects of the shear viscosity due to electron collisions, and the mutual friction processes arising from the scattering of the electrons off the
magnetic fields entrapped in the core of the superfluid neutron vortices have been taken into account  \cite{Lindblom:1999wi,Lee:2002fp, Haskell:2009fz}.
In this article we only analyze the effect of the shear viscosity in the r-mode instability window.  
The superfluid phonons also contribute both to the
bulk viscosities, and to mutual friction, but the microscopic computations necessary to evaluate their effect on the damping of the r-modes
have not been yet carried out.

\begin{figure}[t]
\begin{center}
\includegraphics[width=0.7\textwidth]{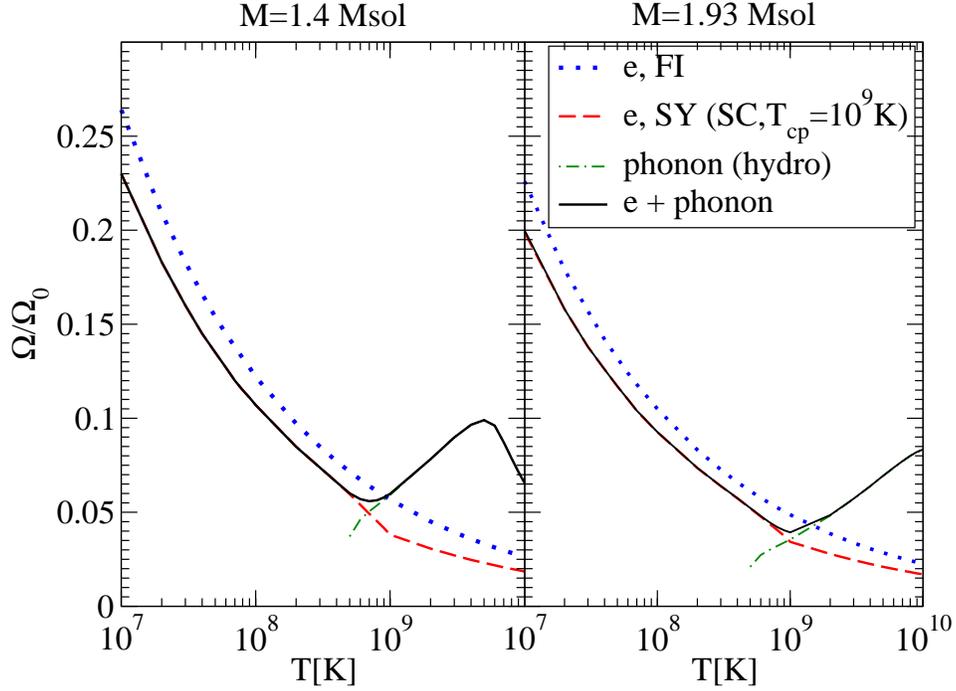}
\end{center}
\caption{(Color online) R-mode instability window for superfluid neutron stars. The critical
(normalized) frequency is plotted for two neutron star mass configurations (1.4 ${\rm M_{sol}}$ and 1.93 ${\rm M_{sol}}$) as a function of temperature for the different dissipative processes studied:  electron collisions assuming static photon screening from Flowers and Itoh (FI) [Eq.~(\ref{ee})] (dotted lines) and due to the inclusion not only of the longitudinal plasmon exchange but also taking into account the transverse plasmon exchange from Shternin and Yakovlev (SY) in proton superconducting matter (SC) with a transition temperature for protons of $T_{cp}=10^9$ K [Eqs.~(\ref{etaee},\ref{mix},\ref{transsup},\ref{longsup})] (dashed lines), the hydrodynamical phonon shear viscosity (dashed-dotted lines), and the sum of both the shear viscosity due to electron collisions from Shternin and Yakovlev and the  hydrodynamical phonon shear viscosity (solid lines). Here $\Omega_0 =\sqrt{G \pi \bar{\rho}}$, where $\bar{\rho}$ is the average mass density of the star.}
\label{fig:freq}
\end{figure}

In this  paper 
%we simply study the impact of the shear viscosity in the r-mode instability window. 
we consider both the shear viscosity as arising from electrons  and phonons.
%, and take the values discussed in Sec.~\ref{shear-sec}. 
Close and below the phase transitions, both for superfluidity and superconductivity,
one might not expect that nucleon-nucleon collisions will result in an exponentially suppressed contribution to  $\eta$. However, we don't take into
account those contributions for the study of r-mode damping, as even in the normal phase it has been shown that
$\eta_e$ dominates.
Thus we only consider electrons and phonons to study the roots of the curve
\be
- \frac{1} { |\tau_{\rm GR} (\Omega) |} +\frac{1}{\tau_{\eta} (T)} = 0  \ ,
\ee
 where the characteristic time scales are given by the expressions of Ref.~\cite{Lindblom:1998wf} (in C.G.S. units)
 \be
\frac{1} { |\tau_{\rm GR} (\Omega) |} = \frac{32 \,\pi \,G\, \Omega^{2l+2}}{c^{2l+3}} \frac{(l-1)^{2l}}{\left( (2l+1)!! \right)^2} \left( \frac{l+2}{l+1} \right)^{2 l+2}  \int^R_0 \rho r^{2 l+2} dr \ ,
\ee  
and 
\be
\frac{1}{\tau_{\eta} (T)}  = (l-1) (2l +1) \int^R_{R_c} \eta r^{2l} dr   \left(\int^R_0 \rho r^{2l+2} dr \right)^{-1}  \ .
\label{ec-hidro-shear}
\ee

We study only  r-modes with  $l = 2$ because they are the dominant ones \cite{Lindblom:1998wf,Andersson:2000mf}. 
For the calculation of the characteristic time associated to the phonon shear viscosity we only take into account the phonon contribution in the hydrodynamical regime, that is, due to  collisions of phonons among themselves. This is achieved by introducing a temperature-dependent lower limit ($R_c$) in the integral for the shear viscosity in Eq.~(\ref{ec-hidro-shear}). The lower the limit is, the higher the density is for which phonons are hydrodynamic. The consideration of only phonon processes within the core results from the fact that this characteristic time is derived from the dissipation due to phonon collisions in the bulk, that is considering dissipations in small volume elements throughout the whole star, and it is not related to the viscous process coming from the interaction of phonons with the crust. This latter contribution, as a surface dissipation process, must be investigated separately. In any case, the dissipation associated with electron collisions takes over whenever the dissipation due to the interactions of phonons with the crust overcomes the one due phonon-phonon collisions, as a first rough estimate indicates in Section \ref{sec-shear-ph}.

In Fig.~\ref{fig:freq} we show the r-mode instability  window of the neutron star as arising from considering only shear viscous damping
for two star mass configurations, 1.4 ${\rm M_{sol}}$ and 1.93 ${\rm M_{sol}}$.  The corresponding radius of those configurations are 11.7 Km and 10.2 Km, respectively, while the central densities are, in turn, 3.2 $n_0$ and 7.6 $n_0$. Let us recall again that we are using the APR EoS in a causal parametrization of Ref.~\cite{Heiselberg:1999mq} as a common benchmark  in order to assess the value of the mass density as a function of r, as well as the values of the shear viscosity. The critical frequency in Fig.~\ref{fig:freq} is given in terms of $\Omega_0 =\sqrt{G \pi \bar{\rho}}$, where $\bar{\rho}$ is the average mass density of the star, and it is  displayed as a function of the temperature for the different processes studied. We plot the old estimate of Flowers and Itoh assuming dominance of  electron collisions with static photon screening [Eq.~(\ref{ee})] (dotted lines). We also display the estimate from Shternin and Yakovlev that considers  the longitudinal and transverse plasmon exchange in superconducting matter with a transition temperature of $T_{cp} \sim 10^9 {\rm K}$ \cite{Shternin:2008es} in [Eqs.~(\ref{etaee},\ref{mix},\ref{transsup},\ref{longsup})]  (dashed lines) and the  hydrodynamical phonon shear viscosity  (dashed-dotted lines). Moreover, we display the sum of both the shear viscosity due to electron collisions from Shternin and Yakovlev and the  hydrodynamical  phonon shear viscosity (solid lines). 
%Due to the fact that we are treating two different fluids, the shear viscosity in this case will be the sum of both contributions or, equivalently, one has to sum the frequencies or the inverse time scales associated to the hydrodynamic modes of the two fluids. 

We observe that for low temperatures $T < 7 \times 10^8 {\rm K}$ for a star of 1.4 ${\rm M_{sol}}$, and  $T < 10^9 {\rm K}$  for a mass configuration of 1.93 ${\rm M_{sol}}$ the electron contribution to the shear viscosity dominates and governs the processes of r-mode damping.
However, the early computations of $\eta_e$ had overestimated the value of $\eta_e$ by few orders of magnitudes, as shown in Fig.~\ref{fig:ee}, which lead also to an underestimation of the r-mode instability window. 
%At very low $T$ the critical frequency turns out to be almost two times smaller than that predicted by the old computation of $\eta$ by Flowers and Itoh.

The dissipation stemming from phonon processes in the hydrodynamical regime is non-negligible for $T  \gtrsim 7 \times 10^8 {\rm K}$ for a star of 1.4 ${\rm M_{sol}}$ while temperatures of $T \gtrsim 10^9 {\rm K}$ are needed for a mass configuration of 1.93 ${\rm M_{sol}}$. This can be understood as follows. A star of 1.4 ${\rm M_{sol}}$ has a central density of $n\sim 3\, n_0$. And, from the middle panel of Fig.~\ref{fig:ee-ph},  we see that for  $T \sim 10^9 {\rm K}$  the processes involving phonons govern the dissipation for densities up to 3 $n_0$. Indeed, as seen in Fig.~\ref{fig:freq}, even for lower temperatures of $T\sim 7 \times 10^8 {\rm K}$ this would be the case. Although for $n_0 \lesssim n \lesssim 3 n_0 $ the shear viscosity is governed by the ballistic regime, the collisions of phonons among themselves start being important for low densities. Thus, for a low-mass star of 1.4 ${\rm M_{sol}}$ and $T\sim 10^9 {\rm K}$, the contribution to the r-mode instability window of the phonon dissipation processes in the hydrodynamical domain are non-negligible.  For a mass configuration of 1.93 ${\rm M_{sol}}$, higher densities are reached at the center of the star of $n \sim (7$-$8)\,n_0$. Then, higher temperatures are needed so that the processes involving hydrodynamic phonons will govern the dissipation (right panel of Fig.~\ref{fig:ee-ph}). In this case, the temperature regime where phonon processes dominate over electron collisions is of $T \gtrsim  10^9 {\rm K}$ . These results can be inferred directly from the solid lines in Fig.~\ref{fig:freq}. From the above considerations we can conclude that collisions among phonons should be taken into account for $T \gtrsim (10^9-10^{10}) \rm{K}$ in order to assess the r-mode instability window.

Note that we are assuming that phonons are well-defined quasiparticles 
 throughout the core of the neutron star for $T \gtrsim (10^9-10^{10}) \rm{K}$, close to $T_c$. As mentioned in the Introduction, the possible overdamping and disappearance of phonons has been studied in \cite{Leggett:1966zz, Leinson:2010ru}. Leggett \cite{Leggett:1966zz} found the conditions for the existence of phonons at finite $T$, relating their damping to the sign and value of the $F_0$ Landau parameter of the Fermi liquid. Recent calculations of the $F_0$ Landau parameter in pure neutron matter \cite{Benhar:2012jk,Holt:2012yv} seem to indicate that phonons could only survive for $n \gtrsim 1.5-2 n_0$  for temperatures around $T=0$ or $T_c$, whereas the repulsive character of the nuclear interactions at high densities implies undamped phonons in the interior of high-mass neutron stars for all $T \lesssim T_c$. A more detailed analysis in $\beta$-stable matter is, however, needed.  In our specific case, for the use of the APR EoS, it is not feasible to check the Leggett criteria for the existence of superfluid phonons, as we do not have access to the corresponding nucleon-nucleon effective interactions in order to extract the Landau parameters. Thus, we can only rely on these previous works and assume that phonons are not damped at high temperatures around $T_c$ in the interior of the high-mass neutron star, where the phonon contribution to the shear viscosity and, hence, to the r-mode damping turns out to be most important.  For low-mass neutron stars of 1.4 ${\rm M_{sol}}$ with densities in the core  around $ 3 n_0$, an overestimate of the phonon viscous processes to the r-mode damping could be expected.
 
%The dissipation stemming from phonon processes  takes over for $T \sim 5 \times 10^8 {\rm K}$ for a star of 1.4 ${\rm M_{sol}}$ while temperatures of $T \sim 10^9 {\rm K}$ are needed for a mass configuration of 1.93 ${\rm M_{sol}}$. This can be understood as follows. A star of  1.4 ${\rm M_{sol}}$ has a central density of $n\sim 3\, n_0$. And, from the middle panel of Fig.~\ref{fig:ee-ph},  we see that for  $T \sim 10^9 {\rm K}$  the processes involving phonons govern the dissipation for densities up to 3 $n_0$. Thus, for a low-mass star of 1.4 ${\rm M_{sol}}$ and $T\sim 10^9 {\rm K}$, the phonon shear viscosity will be the dominant dissipative process. Indeed, as seen in Fig.~\ref{fig:freq}, even for lower temperatures of $T\sim 5 \times 10^8 {\rm K}$ this would be the case. For a mass configuration of 1.93 ${\rm M_{sol}}$, higher densities are reached at the center of the star of $n \sim (7$-$8 )\,n_0$. Then, higher temperatures are needed so that the processes involving phonons will govern the dissipation (middle and right panel of Fig.~\ref{fig:ee-ph}). In this case, the temperature regime where phonon processes dominate over electron collisions is of $T \gtrsim 10^9 {\rm K}$ . These results can be inferred directly from the solid lines in Fig.~\ref{fig:freq}. From the above considerations we can conclude that phonon processes should be taken into account for $T \gtrsim (10^8-10^9) \rm{K}$ in order to assess the r-mode instability window.

%%%%%%%%%%%%%%%%%%%%%%%%%%%%%%%%%%%%%%%%%%%%%%%%%%%%%%%%%%%%%% 
\section{Conclusions}
\label{conclu} 
%%%%%%%%%%%%%%%%%%%%%%%%%%%%%%%%%%%%%%%%%%%%%%%%%%%%%%%%%%%%%

In this article we have made a first attempt to assess whether the dissipation processes arising from the scattering of superfluid phonons
can have any impact on the r-mode instability window associated to neutron stars with superfluid and superconducting cores.
We have so far only considered how shear viscous damping in the core of the star affects the r-mode instability window, 
leaving for a future analysis a much more complete study. Up to now the r-mode instability window of superfluid neutron stars
had been analyzed taking into account only the shear viscosity due to electron collisions and the mutual friction processes arising from the scattering of the electrons off the
magnetic fields entrapped in the core of the superfluid neutron vortices, or the viscous Eckman layer effect. The superfluid phonons
should also contribute to mutual friction, through their
scattering with rotational vortices of the star (a computation of this effect in the ballistic regime of phonons was carried out for the CFL phase in
 Ref.~\cite{Mannarelli:2008je}), or to the bulk viscosity coefficients. But the microscopic computations necessary for the evaluation of their impact on  the r-mode damping
 have not yet been carried out. 

We have considered a simplified model of a neutron star made up of neutrons, protons and electrons, described by a causal parametrization of the APR EoS as a common benchmark.
The star is composed by a rigid crust, and a superfluid and superconducting core. While there are solid theoretical arguments to believe that
neutron and proton pairing occur in the core of the star, the precise values of the different gaps are very model dependent.  We assume a critical
temperature for the onset of neutron superfluidity in the range of $T_c \sim 10^{10}$ K while a transition temperature for proton superconductivity of $T_c \sim 10^9$ K is used \cite{Shternin:2008es}.

Our computations for the shear viscosity as arising from superfluid
phonons assume that superfluidity occurs in the whole core of the star. As the phonon collisions are dominant for $T \gtrsim  10^9$ K, we are indeed assuming that the neutron pairing gap should be $\Delta \sim 2 T_c \gtrsim 0.2$ MeV.

%Our computations for the shear viscosity as arising from superfluid
%phonons do not depend on the value of the neutron gap, and then, we don't need to rely on any specific model for neutron pairing. We simply assume
%that superfluidity occurs in the whole core of the star.
%We consider one specific model for proton pairing, as the electron viscosity does depend on the values of the proton gap, following the results
%of Ref.~\cite{Shternin:2008es}.
%In this way, the values  for the electron viscosity presented in Fig.~\ref{fig:ee} are thus model dependent.
%However, given the similar values for the proton gap in \cite{elgaroynpa96, elgaroyprl76,elgaroyprl77}, we expect that our results for the shear viscosity and the r-mode instability window will be not much affected by  the model of proton pairing.

We have considered two neutron star mass configurations, of 1.4 ${\rm M_{sol}}$ and 1.93 ${\rm M_{sol}}$. The radius and central densities
of these two sort of stars are quite different. Our analysis suggests that at very low $T$ the electron shear viscosity dominates over the phonon viscosity when
obtaining the r-mode instability window. Previous studies of the r-mode instability window were based on a computation of $\eta_e$ that was overestimating its value by few orders of magnitude, which results in an underestimation of the r-mode instability window. The role of superfluid phonons in the hydrodynamical regime starts to be relevant only at $T  \gtrsim  7 \times 10^8$ K for the 1.4 ${\rm M_{sol}}$ star, and for $T \gtrsim 10^9$ K for the 1.93 ${\rm M_{sol}}$. This difference can be understood from the fact that the central densities reached in the two cases differ by almost a factor of two. We thus find that superfluid phonons only contribute at relatively high $T$, and their effect might only be needed to study the r-mode evolution of hot young neutron stars.

%We have considered two neutron star mass configurations, of 1.4 ${\rm M_{sol}}$ and 1.93 ${\rm M_{sol}}$. The radius and central densities
%of these two sort of stars are quite different. Our analysis suggests that at very low $T$ the electron shear viscosity dominates over the phonon viscosity when
%obtaining the r-mode instability window. Previous studies of the r-mode instability window were based on a computation of $\eta_e$ that was overestimating
%its value by few orders of magnitude, which results in an underestimation of the r-mode instability window.
%The recent computations of $\eta_e$ of Ref.~\cite{Shternin:2008es} suggests  smaller critical values of the frequency can be up to half of the values that were predicted in the literature for sufficiently low $T$. 
%The role of superfluid phonons starts to be relevant only at $T  \sim 5 \times 10^8$ K for the 1.4 ${\rm M_{sol}}$ star, and for $T \sim 10^9$ K for the  1.93 ${\rm M_{sol}}$.
% This difference can be understood from the fact that the central densities reached in the two cases differ by almost a factor of two. 
% At lower $T$ superfluid phonons in the core are basically in a ballistic regime, and the dissipation associated to their interactions with the crust are not strong enough
 %so as to compete with the dissipation associated with electron collisions.  We thus find that superfluid phonons only contribute at relatively high $T$, and their effect might only be needed to study the r-mode evolution of hot young neutron stars.

Since we find a serious modification of the r-mode instability window for superfluid neutron stars as arising only from the shear viscous damping,
the Eckman layer damping time scale for the r-modes should be also modified. This is so because this effect, which measures the rubbing effect between the core and the crust
of the star, also depends on the value of the shear viscosity in the core of the star. This analysis, together with a more complete study of all the dissipation mechanisms
associated to superfluid phonons, will be the subject of future studies.

 \acknowledgments{We thank S. Reddy for discussions on the  superfluid phonon-electron interactions and J.A. Oller for  discussions on the effective field theory for nucleon-nucleon interaction in dense matter.
 We also specially thank M.~Mannarelli for a critical reading of the manuscript, and for previous collaborations in related matters.
 This research was supported by Ministerio de Ciencia e Innovaci\'on under contract FPA2010-16963. LT acknowledges support from the Ramon y Cajal Research
Programme from Ministerio de Ciencia e Innovaci\'on and from FP7-PEOPLE-2011-CIG under Contract No. PCIG09-GA-2011-291679.

\appendix

\section{Evaluation of the coupling constant of the phonon-electron interactions}
\label{appendixa}

The value of the coupling constants $g$ and $a_s$ of Eq.~(\ref {Lag-phon-el}) can be computed from the microscopic theory if one integrates
out the heavy fields of the theory. Assuming that the nucleons are paired in a s-wave channel, the computation can be easily done
in the low density domain.

Let us start by recalling the microscopic theory that describes the nucleon dynamics when the momenta of the nucleons is small, i.e., 
smaller than the pion mass. The basic Lagrangian for the nucleon interactions at low momentum
can be described by the effective field theory \cite{Weinberg:1990rz}
\begin{eqnarray}
{\cal L}_N &= & 
 \psi_n^\dagger ( i \partial_t - \frac{\nabla^2}{2 m} + \mu_n) \psi_n + \psi_p^\dagger \left( i (\partial_t + i A_0) - \frac{(\nabla + i {\bf A})^2}{2 m} + \mu_p \right) \psi_p - \frac{C^{nn}_0}{2} \left( \psi_n^\dagger(x) \psi_n(x)\right) \left( \psi_n^\dagger(x) \psi_n(x)\right) 
\nonumber
 \\
 &-& \frac{C^{pp}_0}{2} \left( \psi_p^\dagger(x) \psi_p(x)\right) \left( \psi_p^\dagger(x) \psi_p(x)\right) 
-  \frac{C^{np}_0}{2 }\left( \psi_n^\dagger(x) \psi_n(x)\right) \left( \psi_p^\dagger(x) \psi_p(x) \right) + \cdots \ ,
\end{eqnarray}
where $\psi_n,\psi_p$ are the neutron and proton wave-functions, $A_\mu$ is the electromagnetic gauge potential, and  $C_0^{ij}$ are coupling constants that can be related to the different nucleon-nucleon  scattering lengths $a_{ij}$, 
when these are small,  $ C^{ij}_0= 4 \pi a_{ij}/m$. Note that we have taken into account isospin breaking terms in the contact interactions, and that we have not  displayed 
other contact interactions in a higher spin channel.

It is convenient to write down the neutron wavefunction as
\begin{equation}
\psi_n = \psi_{n,0} e^{i \varphi} \ ,
\end{equation}
where the phase of the neutron wavefunction is the Goldstone field.
Then it is easy to check that the kinetic term associated to $\psi_{n,0}$ is the same as that for the whole neutron wavefunction, where one only
has to substitute the standard derivatives by covariant derivatives
\begin{equation}
 \psi_n^\dagger ( i \partial_t - \frac{\nabla^2}{2 m} + \mu_n) \psi_n =
 \psi_{n,0}^\dagger ( i (\partial_t + i {\tilde A}_0) - \frac{(\nabla + i {\bf \tilde A})^2}{2 m} + \mu_n) \psi_{n,0} \ ,
 \end{equation}
where ${\tilde A}_\mu = (\partial_t \varphi, \nabla \varphi)$.
Integrating out the field  $\psi_{n,0}$, we obtain the   Lagrangian for the superfluid phonon, see Eq.~(\ref{LO-Lagran}).
The value of the coupling constants  $g$ and $a_s$  can be obtained after computing the diagram displayed in Fig.~\ref{fig:diag}.
For the purpose of the present discussion, it will be enough to present a quick estimate of the coupling constant $g$. We do not calculate
the value of $a_s$, as it turns out that it may not affect in any relevant way our conclusions about  the phonon-electron interactions in the core of the star.
The superfluid phonon momentum in the star is of the order of  $T/c_s$,  where $c_s$ is the value of the speed of sound,
and it is much less than the  Fermi momentum of the neutrons
or protons in the star. Thus, one can evaluate the neutron and proton loops of the diagram of Fig.~\ref{fig:diag} by neglecting the phonon momentum in the loop.
Every nucleon loop contributes as $(m\  p_n)/\pi^2$, where $p_n$ is the Fermi momentum of the nucleon 
(see for example Ref.~\cite{Schafer:2006yf}). The photon propagator is also assumed to be dominated by the value of the
Debye or Meissner photon mass. Thus, one concludes that 
\be
g  \approx  \frac{2 \,p_{n} m \,c_s}{\pi \sqrt{\rho_n}} a_{np} \ .
\ee

\end{document}